\documentclass[prd,aps,twocolumn,floatfix,showpacs,nofootinbib]{revtex4-1}

\usepackage{graphicx,psfrag}
\usepackage{mathrsfs}
\usepackage{amsmath,amsfonts,amssymb,bm}
\usepackage{multirow,enumerate}
\usepackage{comment}
\usepackage[hidelinks]{hyperref}
\usepackage{color,xspace}
\usepackage{xcolor}
\usepackage[utf8]{inputenc}  
\usepackage{lineno}

\newcommand{\be}{\begin{equation}}
\newcommand{\ee}{\end{equation}}
\newcommand{\bea}{\begin{eqnarray}}
\newcommand{\eea}{\end{eqnarray}}
\newcommand{\bel}{\begin{align}}
\newcommand{\eel}{\end{align}}

\def\core{{\texttt{CoRe}}\xspace}
\def\BAM{{\texttt{BAM}}\xspace}
\def\THC{{\texttt{THC}}\xspace}
\def\lorene{{\texttt{Lorene}}\xspace}
\def\SGRID{{\texttt{SGRID}}\xspace}

\def\GMc2{{\rm G M_{\odot} c^{-2}}}

\usepackage{color}

\definecolor{cyan}{rgb}{0,0.9,0.9}
\definecolor{orange}{rgb}{0.9,0.5,0}
\definecolor{magenta}{rgb}{1,0,1}
\definecolor{purple}{rgb}{0.8,0.4,0.8}
\definecolor{gray}{rgb}{0.5,0.5,0.5}
\definecolor{proposal}{rgb}{0.6,0.8,1}
\definecolor{burgundy}{rgb}{0.5, 0.0, 0.13}

\begin{document}

\title{\texttt{CoRe} database of binary neutron star merger waveforms and 
       its application in waveform development}

\title{\texttt{CoRe} database of binary neutron star merger waveforms}

\author{Tim  Dietrich$^{1,2}$, 
David  Radice$^{3,4}$,
Sebastiano  Bernuzzi$^{5,6}$,
Francesco  Zappa$^{2,7}$,
Albino  Perego$^{6,8}$,
Bernd  Br\"ugmann$^5$,
Swami Vivekanandji Chaurasia$^5$, 
Reetika Dudi$^5$,
Wolfgang  Tichy$^9$,
Maximiliano  Ujevic$^{10}$}

\affiliation{${}^1$ Nikhef, Science Park, 1098 XG Amsterdam, The Netherlands}
\affiliation{${}^2$ Max Planck Institute for Gravitational Physics (Albert Einstein Institute), Am M\"uhlenberg 1, Potsdam 14476, Germany}
\affiliation{${}^3$ Institute for Advanced Study, 1 Einstein Drive, Princeton, NJ 08540, USA}
\affiliation{${}^4$ Department of Astrophysical Sciences, Princeton University, 4 Ivy Lane, Princeton, NJ 08544, USA}
\affiliation{${}^5$ Theoretical Physics Institute, University of Jena, 07743 Jena, Germany}  
\affiliation{${}^6$ Istituto Nazionale di Fisica Nucleare, Sezione Milano Bicocca, gruppo collegato di Parma, I-43124 Parma, Italy}  
\affiliation{${}^7$ Dipartimento di Scienze Matematiche Fisiche ed
  Informatiche, Universit\'a di Parma, I-43124 Parma, Italia}
\affiliation{${}^8$ Dipartimento di Fisica, Universit\`{a} degli Studi di Milano Bicocca, Piazza della Scienza 3, 20126 Milano, Italia}
\affiliation{${}^9$ Department of Physics, Florida Atlantic University, Boca Raton, FL 33431 USA}
\affiliation{${}^{10}$ Centro de Ci\^encias Naturais e Humanas, Universidade Federal do ABC,09210-170, Santo Andr\'e, S\~ao Paulo, Brazil}

\date{\today}

\begin{abstract} 
We present the Computational Relativity (\core) collaboration's
public database of gravitational waveforms from binary neutron star
mergers. 
The database currently contains 367 waveforms from numerical
simulations that are consistent with general relativity and that
employ constraint satisfying initial data in
hydrodynamical equilibrium. It spans 164 physically distinct
configuration with different binary parameters (total binary mass, mass-ratio,
initial separation, eccentricity, and stars' spins) and simulated physics.
Waveforms computed at multiple grid resolutions and extraction radii
are provided for controlling numerical uncertainties.
We also release an exemplary set of 18 hybrid waveforms constructed
with a state-of-art effective-one-body model spanning the
frequency band of advanced gravitational-wave detectors. 
We outline present and future applications of the database to
gravitational-wave astronomy.
\end{abstract}

\maketitle

\begin{figure*}[t] 
  \centering
  \href{http://www.computational-relativity.org}{\includegraphics[width=\textwidth]{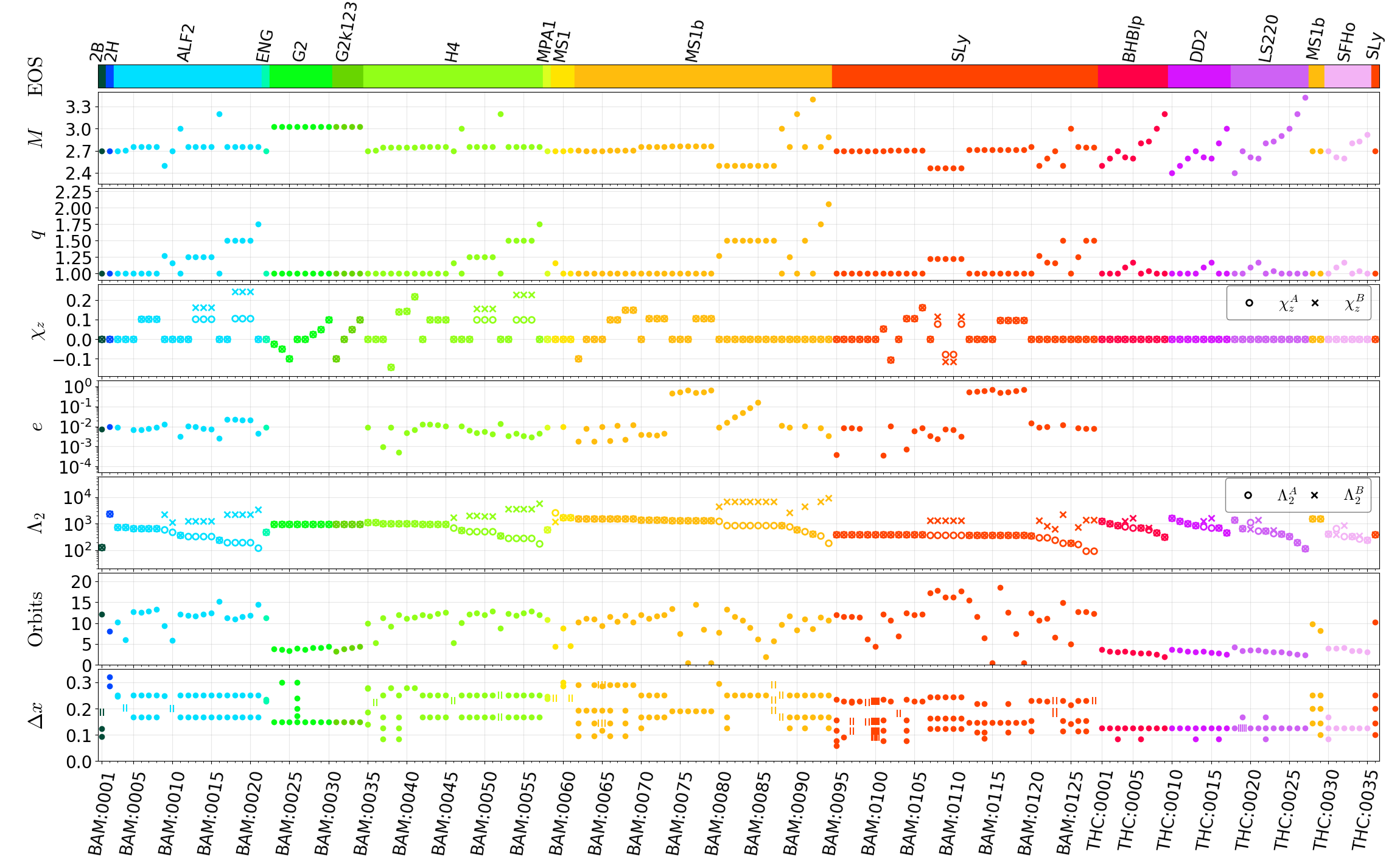} }
  \caption{Simulations contained in the \texttt{CoRe} database. 
  We present the total mass $M$, the mass ratio $q$, the individual dimensionless 
  spins $\chi^{A,B}_{z}$, the eccentricity $e$ 
  [no eccentricity measurement is given for too short simulations], 
  the individual quadrupolar tidal parameters $\Lambda^{A,B}_{2}$, 
  the number of orbits [note that for highly eccentric orbits close to head-on, the 
  number of orbits can drop below $1$], 
  and the employed resolution $\Delta x$ of the finest level covering the
  entire NS for different configurations.
  Different colored markers refer to different EOS, see top color bar. 
  In the last panel we also include 
  simulations with different grid resolutions and 
  numerical methods (fluxes, mesh refinement strategies etc.); 
  simulations of a fixed configuration performed at the same resolutions
  but using different methods are marked with vertical bars in this panel.}
  \label{fig:simulations}
\end{figure*}

The era of gravitational-wave (GW) astronomy 
has been inaugurated with the direct
detection of GWs from binary black hole (BBH)
mergers~\cite{Abbott:2016blz,Abbott:2016nmj,Abbott:2017vtc,Abbott:2017oio,Abbott:2017gyy}
soon followed by the breakthrough observation of GWs 
and electromagnetic (EM) signals from a binary neutron star (BNS) 
collision~\cite{TheLIGOScientific:2017qsa,Monitor:2017mdv,GBM:2017lvd,Coulter:2017wya}.

Numerical relativity (NR) is the fundamental tool to study
GWs from systems in the strong-field regime, and it has
crucially supported the first discoveries. 
In particular, different NR groups have publicly released BBH
simulation data~\cite{Mroue:2013xna,Healy:2017psd,Jani:2016wkt}.
These catalogs have been the cornerstone of many scientific
results. They have been used to improve our understanding of the 
merger
dynamics~\cite{Lousto:2013wta,Tiec:2013twa,Nagar:2015xqa,Varma:2014jxa,Keitel:2016krm,Gerosa:2016vip}, 
to develop waveform models~\cite{Pan:2013rra,Damour:2014sva,
Taracchini:2013rva,Husa:2015iqa,Purrer:2015tud,
Babak:2016tgq,Bohe:2016gbl,Huerta:2016rwp} including 
surrogates~\cite{Field:2013cfa,Blackman:2015pia}, and  
to validate LIGO-Virgo parameter 
estimation pipelines~\cite{Abbott:2016apu,Abbott:2016wiq}. 

Following the first successful BNS merger simulations in full general
relativity~\cite{Shibata:1999wm,Shibata:1999hn}, the NR community has
made tremendous progresses on several aspects of the problem:
%
(i) the exploration of the effect of different equations of state
(EOSs), total mass and mass-ratio on the merger dynamics~\cite{Hotokezaka:2011dh,Bernuzzi:2014kca,Foucart:2015gaa,Bernuzzi:2015opx,Dietrich:2016hky,Lehner:2016lxy,Sekiguchi:2016bjd};
(ii) the development of many-orbits simulations for high-precision GW modeling
\cite{Bernuzzi:2011aq,Hotokezaka:2013mm,Radice:2013hxh,Bernuzzi:2016pie,
Hotokezaka:2015xka,Dietrich:2017aum,Kiuchi:2017pte};
(iii) the exploration of BNS mergers from eccentric orbits and dynamical collisions~\cite{
Gold:2011df,East:2012ww,East:2015vix,East:2016zvv,Radice:2016dwd};
(iv) the inclusion of aligned spins and spin-precession effects 
\cite{Bernuzzi:2013rza,Kastaun:2013mv,Dietrich:2016lyp,East:2015vix,Kastaun:2016elu,Dietrich:2015pxa,Tacik:2015tja,Dietrich:2017xqb}; 
(v) the simulations of magnetic effects in connection to gamma-ray
bursts engines~\cite{Anderson:2008zp,Giacomazzo:2010bx,Rezzolla:2011da, 
Palenzuela:2013hu,Kiuchi:2014hja,Palenzuela:2015dqa,Ruiz:2017due,Kiuchi:2017zzg};
(vi) the study of finite-temperature and composition effects using a
microphysical descriptions of NS matter together with neutrino
transport \cite{Sekiguchi:2011zd,Galeazzi:2013mia,Neilsen:2014hha,
Sekiguchi:2015dma,Palenzuela:2015dqa,Foucart:2016rxm,Foucart:2017mbt};
(vii) the study of mass ejecta and EM counterparts
\cite{Hotokezaka:2012ze,Bauswein:2013yna,Wanajo:2014wha,Lehner:2016lxy,
Radice:2016dwd,Dietrich:2016hky,Dietrich:2016lyp,Foucart:2016rxm,Fujibayashi:2017xsz,
Perego:2017wtu,Hotokezaka:2018aui,Bovard:2017mvn}.
New frontiers in BNS merger simulations are the inclusion of general-relativistic radiation hydrodynamics
\cite{Shibata:2012ty,Foucart:2015vpa,Foucart:2017mbt}
and viscous
hydrodynamics effects~\cite{Radice:2017zta,Shibata:2017jyf,
Shibata:2017xht,Shibata:2017xdx,Fujibayashi:2017puw}. 

Since 2009 our team has contributed to some of the research lines 
mentioned above. Here, we present the largest-to-date public database of 
BNS waveforms composed of new simulations 
and those published in~\cite{Bernuzzi:2013rza,Bernuzzi:2014kca,
Bernuzzi:2014owa,Dietrich:2015iva,Bernuzzi:2015rla,Dietrich:2015pxa,Radice:2016gym,
Radice:2016rys,Bernuzzi:2016pie,Dietrich:2016hky,Dietrich:2016lyp,Radice:2017zta,
Radice:2017lry,Dietrich:2017feu,Dietrich:2017aum,Dietrich:2017xqb,Zappa:2017xba,
Dietrich:2018upm}.
The combined set of simulations required about $150$ million CPU-hours
on supercomputers in Europe and the United States.
We publicly release these data with the goal of supporting researchers
and further developments in the field of GW astronomy (\url{www.computational-relativity.org}). 
We plan to extend the database with waveforms from upcoming
simulations and from other groups/codes. 

This article describes the simulation methods in $3+1$ NR, summarizes the quality
of the computed waveforms and the key parameters that characterize the GW,
and concludes outlining some of the many applications. We use
geometrized units $c=G=1$ and express results in terms of solar masses
($M_\odot=1.9889\times 10^{33}$~g) if not otherwise stated. Conversion
factors to CGS are $[L]=GM_\odot/c^2\simeq1.47670\times 10^{5}$~cm and
$[T]=GM_\odot/c^3\simeq4.92549\times10^{-6}$~s. 


\section*{Simulation Methods}      
\label{sec:methods}

\paragraph*{\textbf{Initial data.}}
Initial data are constructed by solving the Einstein constraint
equations in the conformal thin sandwich
formalism and by imposing hydrodynamical equilibrium for the star
fluid~\cite{Wilson:1995uh,Wilson:1996ty,York:1998hy}. 
The fluid's flow is chosen to be either irrotational \cite{Bonazzola:1998yq}, 
or prescribed according to the constant rotational velocity
formalism~\cite{Tichy:2011gw,Tichy:2012rp}. 
Binaries in quasi-circular orbits are built 
imposing a helical Killing vector \cite{Gourgoulhon:2000nn}, whereas
for eccentric orbits an approximate 
``helliptical'' Killing vector is used \cite{Moldenhauer:2014yaa,Dietrich:2015pxa}.
We use either the public \lorene~\cite{LoreneCode} or the
\SGRID~\cite{Tichy:2009yr,Tichy:2012rp,Dietrich:2015pxa} code.
Both codes use multi-domain pseudo-spectral methods with surface fitting coordinates
\cite{Gourgoulhon:2000nn,Ansorg:2006gd}. 

\paragraph*{\textbf{Evolutions.}}
Dynamical simulations are performed using free-evolution schemes for the Einstein
equations and general relativistic hydrodynamics (GRHD).
For the spacetime, we employ either the
BSSNOK~\cite{Nakamura:1987zz,Shibata:1995we,Baumgarte:1998te} formalism or the 
Z4c formalism~\cite{Bernuzzi:2009ex,Ruiz:2010qj,Weyhausen:2011cg,Cao:2011fu,Hilditch:2012fp}. 
The latter has improved constraint propagation and
damping properties with respect to BSSNOK, especially in matter simulations \cite{Bernuzzi:2009ex,Hilditch:2012fp}. 
We use the moving puncture gauge~\cite{Bona:1994a,Alcubierre:2002kk,vanMeter:2006vi,
Campanelli:2005dd,Baker:2005vv}, which can handle automatically
the gravitational collapse without the need
for
excision~\cite{Baiotti:2007np,Thierfelder:2010dv,Dietrich:2014wja}. GRHD
is solved in flux-conservative form \cite{Banyuls:1997zz}. 
Some mergers are simulated with microphysical EOS
and neutrino cooling is taken into account with a leakage
scheme. Viscous effects in GR are also simulated in a few cases using
the large eddy scheme (GRLES) developed in~\cite{Radice:2017zta}. 

We use two different NR codes: \BAM~\cite{Bruegmann:2003aw,Brugmann:2008zz,Thierfelder:2011yi} 
and \THC~\cite{Radice:2012cu,Radice:2013hxh,Radice:2013xpa,Radice:2015nva}.
Both codes use a simple mesh refinement scheme whereby
the grid hierarchy is composed of nested Cartesian boxes, some of which 
can be moved to track the orbital motion of the
stars~\cite{Berger:1984zza,Schnetter:2003rb,Brugmann:2008zz}. 
The grid setup is controlled by the resolution $\Delta x$ in the
finest levels. The finest refinement levels cover entirely the NSs during the
inspiral. The other levels are constructed by progressively coarsening the
resolution by factors of two and extend to the wave-extraction zone.
Discretization is based on fourth (or higher) order
finite-differencing stencils and GRHD is handled with either standard finite-volume or 
high-order finite-differencing high-resolution shock-capturing
methods~\cite{Radice:2013hxh,Bernuzzi:2016pie}. 
The \THC code also implements a neutrino leakage scheme and the
GRLES~\cite{Radice:2016dwd,Radice:2017zta}. 

\paragraph*{\textbf{Wave extraction.}} 
GWs are extracted on coordinate spheres with radius $r$ using the
spin-weighted $s=-2$~spherical harmonics decomposition of the Weyl scalar
$\Psi_{4}$, e.g.,~\cite{Brugmann:2008zz}. Some of the \THC simulations
employ the Cauchy characteristic extraction technique to obtain $\Psi_{4}$
at future null infinity \cite{Reisswig:2009us}. The metric
multipoles are reconstructed using the fixed frequency method
\citep{Reisswig:2010di}.
We release the $\ell=m=2$ metric multipole $h_{22}$ as a function of the coordinate time
$t$ and of the retarded time $u=t-r_*$, where $r_*(r)$ is the 
tortoise coordinate defined by assuming $r$ is the
isotropic radius and using the binary total mass for the Schwarzschild
spacetime.
We also release the GW energy and angular
momentum emitted during the simulation that are computed as 
in \cite{Damour:2011fu,Bernuzzi:2012ci}.
Our waveforms are extracted at different extraction radii and can be further 
extrapolated to obtain null-infinity estimates,
e.g.~\cite{Scheel:2008rj, Reisswig:2009rx,Bernuzzi:2011aq}. 


\section*{Input physics}
BNS simulations require several assumptions on the NS fluid and input
models describing the matter interactions.
The yet unknown EOS is among the most important quantities 
determining the NS properties and the binary dynamics.
It determines the tidal deformations and interactions of 
the stars during the
inspiral~\cite{Damour:1983a,Hinderer:2007mb,Damour:2009kr,Hinderer:2009ca,Damour:2012yf},  
the lifetime and rotation frequency of the merger 
remnant~\cite{Bauswein:2011tp,Hotokezaka:2013iia,Takami:2014zpa,Bernuzzi:2015opx,Radice:2016rys}, 
and the amount of unbound matter ejected during the merger 
process,
e.g.~\cite{Hotokezaka:2012ze,Bauswein:2013yna,Dietrich:2015iva,Lehner:2016lxy}. 
We release data for 16 different EOSs.
Two EOSs are polytropic models with adiabatic index $\Gamma=2$
\cite{Bernuzzi:2013rza}.
Nine EOSs are zero-temperature nuclear physics model represented by 
piecewise polytropic fits~\cite{Read:2008iy}. They are augmented with a
$\Gamma$-law pressure component 
during the simulation to approximate temperature effects
\cite{Shibata:2005ss}. Five EOSs are tabulated finite-temperature
microphysical models developed 
in~\cite{Lattimer:1991nc,Typel:2009sy,Hempel:2009mc,Steiner:2012rk,Banik:2014qja},
which we also release.
Finite-temperature effects are crucial during and after merger, when
compressional and shock heating are present,
e.g.~\cite{Bauswein:2010dn,Sekiguchi:2011zd,Kaplan:2013wra,Sekiguchi:2015dma,Lehner:2016lxy}. 

The role of magnetic fields on the post-merger dynamics is
currently a key open question
\cite{Giacomazzo:2010bx,Kawamura:2016nmk,Kiuchi:2017zzg,Radice:2017zta,
Shibata:2017xht,Fujibayashi:2017puw,Radice:2018xqa}. Large
magnetic field instabilities might cause turbulence and induce
viscosity, potentially affecting the merger remnant, mass outflows and the GW emission.
Also, while not directly relevant for GW emission on the dynamical timescale
of our simulations, neutrino transport plays a crucial role in the merger remnant,
e.g.~\cite{Sekiguchi:2011zd,Galeazzi:2013mia,Sekiguchi:2015dma,Foucart:2015gaa,
Palenzuela:2015dqa,Lehner:2016lxy,Radice:2016dwd,Bovard:2017mvn,Perego:2017wtu}. 
We plan to include more data from simulations with advanced radiation transport
schemes and magnetic field effects as robust NR results become available.


\section*{Waveform parameters}      
\label{sec:database}

\begin{figure}[t]
\centering
\href{http://www.computational-relativity.org}{\includegraphics[width=.5\textwidth]{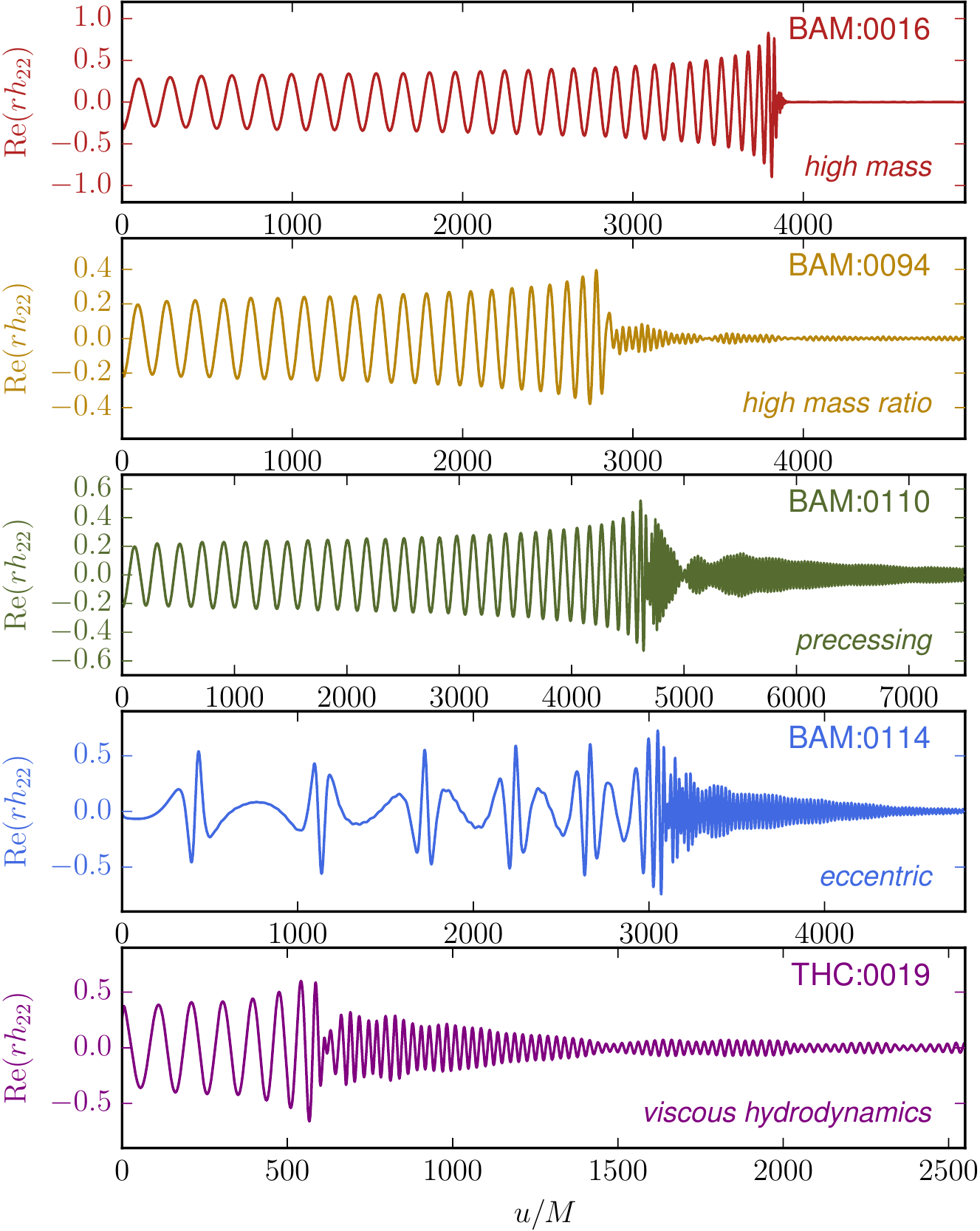} }
\caption{Waveforms from the database showing, from top to bottom panel, the influence of
total mass, mass ratio, spins, eccentricity, and EOS.}
\label{fig:waveforms}
\end{figure}

Figure~\ref{fig:simulations} summarizes our database in terms of the
main parameters that characterize the GWs.

\paragraph*{\textbf{Binary mass.}} 
In contrast to BBHs, BNS dynamics cannot be rescaled by the
binary total mass ($A,B$ label the NSs) 
\begin{equation}
 M = M_A +M_B \ ,
 \label{eq:Mtotal}
\end{equation}
since $M$ enters the description of tidal interactions during 
the inspiral and determines the merger remnant. 

Formation scenarios and the constraints from GW170817 indicate
that NS masses lie within 
$\sim 1.0 {-} 2.3 M_\odot$~\cite{Lattimer:2012nd,ozel:2016oaf,Rezzolla:2017aly,
Shibata:2017xdx,Ruiz:2017due,Margalit:2017dij}. 
Current observations range from 
$\sim 1.0 M_\odot$~\cite{Rawls:2011jw,Ozel:2012ax} 
to $\sim 2.0 M_\odot$~\cite{Demorest:2010bx,Antoniadis:2013pzd} 
or possibly even $\sim 2.3 M_\odot$~\cite{Linares:2018ppq},
with BNS masses varying in $\sim 2.5-2.9$ $M_\odot$~\cite{Latimer:web,Huang:2018ese}.
The wide mass range in our database fully covers the observational 
and a large fraction of the theoretical limits.

GWs from a $M=3.2M_\odot$ merger are shown in the top panel of Fig.~\ref{fig:waveforms}. 
High-mass mergers likely result in a prompt BH formation; while
high-mass BNS emit strong GWs, they are EM 
faint due to smaller ejecta and disk 
masses~\cite{Bauswein:2013yna,Hotokezaka:2013iia}.

\paragraph*{\textbf{Mass ratio.}} 
The mass ratio 
\begin{equation}\label{eq:q}
q = \frac{M_A}{M_B} \geq 1\ ,  
\end{equation}
has a clear imprint on the GW/EM signals: BNS with larger $q$ are less 
luminous in GWs~\cite{Hinderer:2009ca,Dietrich:2016lyp,Zappa:2017xba},
but their larger mass ejecta can power bright EM
transients~\cite{Dietrich:2016fpt,Dietrich:2016lyp,Lehner:2016lxy}.
The isolated NS mass distribution implies mass ratios up to 
$q_{\max} \simeq 2.3$, but population synthesis models predicts
lower values $q_{\max} \simeq 1.8 {-} 1.9$~\cite{Dominik:2012kk,Dietrich:2015pxa}.
The largest observed mass ratio in BNSs is
$q\sim 1.3$~\cite{Martinez:2015mya,Lazarus:2016hfu}.
The \core database contains data with mass ratios up to $q=2.1$, which
is the largest simulated so far~\cite{Dietrich:2015pxa,Dietrich:2016hky}.
In this simulation the companion NS is tidally disrupted during the
merger leading to postmerger GWs with small amplitude 
(second panel in Fig.~\ref{fig:waveforms}).

\paragraph*{\textbf{Spins.}}
The dimensionless spin of a NS in a binary can be defined as 
\begin{equation}
 {\bm\chi}_{A} = \frac{\mathbf{S}_{A}}{M_{A}^2}\ ,
\end{equation}
where the angular momentum $\mathbf{S}_{A}$ is computed 
from the isolated NS with the same 
EOS, rotational velocity, and baryonic mass as the constituents of the
binary~\cite{Bernuzzi:2013rza,Dietrich:2015pxa,Dietrich:2016lyp}.
The maximum NS spin is not precisely known, since it depends on
the EOS, but existing EOS models predict breakup spins below 
$|{\bm\chi}|\sim0.7$, corresponding to spin periods of less than 1
ms~\cite{Lo:2010bj}. The fastest spinning NS in a BNS system is PSR
J1946+2052~\cite{Stovall:2018ouw} which will have $|{\bm\chi}|\sim 0.05$ at merger. 

For spins parallel to orbital angular momentum (say $z$-direction),
the effective spin~\cite{Buonanno:2002fy}  
\begin{equation}
\chi_{\rm eff} = \frac{M_A}{M} \chi^A_z + \frac{M_B}{M} \chi^B_z -
\frac{38}{113} \frac{M_A M_B}{M^2}(\chi^A_z + \chi^B_z) \ ,
\label{eq:chieff}
\end{equation}
is the quantity determining the leading-order spin-orbit effects 
on the phase evolution of the binary.
Spin-orbit interactions quantitatively change the inspiral-merger and
remnant dynamics
~\cite{Damour:2001tu,Campanelli:2006fg,Bernuzzi:2013rza}. Neglecting
their effect can bias the GW parameter
estimation~\cite{Harry:2018hke,Agathos:2015uaa,Dietrich:2018uni}.  
Spin-precession effects in BNS have been first
simulated in Refs.~\cite{Dietrich:2015pxa,Dietrich:2017xqb}; 
the computed GW signal is shown in Fig.~\ref{fig:waveforms} 
(third panel).

\paragraph*{\textbf{Eccentricity.}}
The emission of GWs causes field binaries to circularize to eccentricities
$e \lesssim 10^{-5}$ by the time they enter the LIGO-Virgo band~\cite{Peters:1964zz}. 
Therefore, for an accurate modeling of the GWs, 
it is important to simulate small eccentricity binaries. The residual
(numerical) eccentricity of the initial data can be reduced to $e \lesssim 10^{-3}-10^{-4}$
using an iterative procedure~\cite{Pfeiffer:2007yz,Kyutoku:2014yba,Dietrich:2015pxa}.

On the contrary, dynamically assembled BNSs or those belonging to
hierarchical triplets could be highly eccentric even at the time of
merger~\cite{Bonetti:2018nwo,Lee:2009ca}. 
An example of a highly eccentric merger with $e=0.6$ is shown in
Fig.~\ref{fig:waveforms} (forth panel).
The bursts in the GW amplitude are caused
by the close encounters of the two stars. These encounters also induce 
$f$-mode oscillations which allow an independent constraint of the EOS 
for upcoming 3rd generation GW detectors~\cite{Gold:2011df,Chirenti:2016xys,Chaurasiainprep}. 

\paragraph*{\textbf{Tidal parameters.}}
Tidal interactions in the post-Newtonian (PN) 
formalism are described by a multipolar set of parameters
proportional to the relativistic Love
numbers~\cite{Damour:1983a,Hinderer:2007mb,Damour:2009kr,Binnington:2009bb}. 
The dominant effect depends on the gravitoelectric quadrupolar Love
numbers $k^{A}_2$ and the compactness $C_A$ of the NS through the
expression $\Lambda_2^A=2k_2/(3 C_A^5)$.
Tidal interactions are attractive and enter at leading PN 
order in the GW phasing evolution through the
combination~\cite{Flanagan:2007ix,Damour:2009vw,Damour:2012yf,Favata:2013rwa},  
\begin{equation}
\tilde{\Lambda} = \frac{16}{13} \left[
\frac{(M_A+12M_B)M_A^4}{(M_A+M_B)^5}\Lambda_2^A +
 (A \leftrightarrow B)
\right]  \ .
\end{equation}
The tidal parameter $\tilde{\Lambda}$ is a key quantity to
characterize the non-perturbative regime of the merger dynamics as
shown in~\cite{Read:2013zra,Bernuzzi:2014kca,Zappa:2017xba}
and discussed below. Furthermore, it provides a simple but effective
parameterization of the characteristic GW post-merger
frequencies~\cite{Bernuzzi:2015rla,Rezzolla:2016nxn,Lehner:2016lxy}
and of the disk mass~\cite{Radice:2017lry}.
 
The value of $\tilde{\Lambda}$ for GW170817 is constrained to be
$\lesssim 630$ on the basis of the analysis of the GW signal 
alone~\cite{TheLIGOScientific:2017qsa,Abbott:2018wiz}.
In addition, Refs.~\cite{Radice:2017lry,Coughlin:2018miv} suggested that 
the observation of an EM counterpart to GW170817 allows to place a lower bound on the 
tidal deformability of $\tilde{\Lambda} \gtrsim 400$, Ref.~\cite{Radice:2017lry},
or $\tilde{\Lambda} \gtrsim 200$, Ref.~\cite{Coughlin:2018miv}. 
Further constraints arise from the theoretical modeling of 
matter near nuclear density, e.g.~\cite{SchaffnerBielich:2008kb}, other
astrophysical observations, e.g.~\cite{Ozel:2009da,Neuhauser:2011jg,
Steiner:2012xt,ozel:2016oaf,Watts:2016uzu}, and from the combination of all 
these constraints by considering a large set of possible 
nuclear physics EOSs, e.g.~\cite{Annala:2017llu,Most:2018hfd}.


\section*{Data quality}      

Waveforms' error budgets based on convergence tests and finite radius
extraction have been presented in
\cite{Bernuzzi:2011aq,Radice:2013hxh,Radice:2013xpa,Radice:2015nva,Bernuzzi:2016pie,Dietrich:2017aum,Dietrich:2018upm}. Phase 
convergence is typically observed for about $10-15$ orbits at sufficiently high resolutions,
corresponding to about $\gtrsim96$ grid points per NS
diameter. The error due to finite-radius extraction dominates in the early part of
the simulations, but truncation errors increase towards merger and
afterwards where the uncertainty is the largest
\cite{Bernuzzi:2011aq}. Typical accumulated phase errors up to merger are
estimated as $\delta\phi\sim0.2-1.5$~rad for simulations in which
convergence can be proven. We stress, however, that the lowest
resolutions employed in our runs are not convergent and do not give
quantitatively reliable results for multiple orbits. Post-merger
GWs are typically less accurate, 
but monotonic behavior with grid resolution can be observed at
sufficiently high resolutions, e.g.~\cite{Radice:2016rys}. Our
post-merger data are sufficiently robust to infer the energy and
frequency content, e.g.~\cite{Bernuzzi:2015rla,Radice:2016rys,Zappa:2017xba}.

We assessed systematic errors due to the use of nonlinear
numerical schemes used for GRHD~\cite{Bernuzzi:2012ci,Radice:2013xpa}
with extensive testing of different algorithms and/or extensive code
comparisons. We have tested consistency between \BAM and \THC for 
datasets: 
\texttt{BAM:0097} and \texttt{THC:0036}, 
\texttt{BAM:0063} and \texttt{THC:0029}, 
\texttt{BAM:0064} and \texttt{THC:0028}, 
using exactly the same initial data.  We found that phase
differences are below the estimated uncertainties. 
A simple polytropic EOS setup has also been compared to 
results obtained with the \texttt{SpEC} 
code~\cite{Scheel:2006gg,Szilagyi:2009qz,Buchman:2012dw} with
similar results~\cite{Haas:2016cop}.

We stress that all of our waveforms are computed using constraint-satisfying
initial data in hydrostatic equilibrium. Constraint violating and/or
non-hydrostatic initial data exhibits large unphysical fluid oscillations that
contaminate the GW signals. These oscillations are significantly reduced and
converge to zero if equilibrium is imposed~\cite{Dietrich:2018upm}.  
Systematic errors generated by the initial data were studied
by comparing the evolution of a binary 
produced by \SGRID and \lorene using the same evolution setup
(\texttt{BAM:0026}, \texttt{BAM:0027})~\cite{Bernuzzi:2013rza}.
Differences in the GW phase and collapse time to black-hole were found
to be compatible with those expected from finite grid resolutions effects.


\section*{Applications} 
\label{sec:waveform}

The \core waveform database has wide applicability to the study of
strong-field BNS dynamics and for GW astronomy.  


Our simulations showed that, despite the complexity of
the physics involved, the main quantities characterizing the merger dynamics, like the
mass-rescaled GW frequency and the binding energy per unit mass, are
determined by parameters like $\tilde{\Lambda}$, emerging from 
perturbative (PN and effective-one-body, EOB) analysis~\cite{Bernuzzi:2014kca,Bernuzzi:2015rla}.
About 100 simulations of the \core database were 
used to compute the total GW luminosity in terms of 
tidal parameters and the mass-ratio for {\it all} BNS with aligned
spins $|\chi_z|\lesssim0.14$,
and to set upper limits to the total emitted
energy \cite{Zappa:2017xba}.

A related application is the study of the merger outcome. NR data are 
crucial to understand the formation of massive NS remnant   
\cite{Giacomazzo:2013uua,Foucart:2015gaa,Radice:2018xqa} and prompt black hole
formation at merger~\cite{Kiuchi:2010ze,Bauswein:2013jpa,Hotokezaka:2013iia}. 

The data we provide can be used to verify and develop inspiral-merger
waveform models for LIGO-Virgo analysis. \BAM simulations have 
been already used in the development of the \texttt{TEOBResum} model
\cite{Bernuzzi:2014owa}. Further analytical-numerical comparisons showed that 
state-of-art tidal EOB models might underestimate tidal effects at
merger for stiff EOS and small $M$ \cite{Dietrich:2017feu}.
Our \textit{spinning} BNS are currently used to test the
performances of the \texttt{TEOBResumS} model~\cite{Damour:2014sva,Nagarinprep}. 

High-resolution \BAM simulations (\texttt{BAM:0037},
\texttt{BAM:0064}, \texttt{BAM:0095}) were employed to construct the tidal phase model
\texttt{NRtidal}~\cite{Dietrich:2017aum,Dietrich:2018upm,Dietrich:2018uni}.  
The latter is a closed-form expression fitting the inspiral-merger GW 
composed of PN, \texttt{TEOBResum}, and NR data used to
augment any BBH waveform model 
with tidal effects \cite{Dietrich:2018uni}. Notably,
\texttt{NRtidal} was used in the LIGO-Virgo analysis of
GW170817~\cite{TheLIGOScientific:2017qsa,Chatziioannou:2018vzf,Abbott:2018wiz,Abbott:2018exr}, and
other groups are using similar approaches for GW modeling \cite{Kawaguchi:2018gvj}.

A main open challenge is the modeling of GWs 
from merger remnants \cite{Hotokezaka:2013iia,Bauswein:2012ya,
Bauswein:2014qla,Takami:2014zpa,Takami:2014tva,Bauswein:2015yca,
Clark:2015zxa,Rezzolla:2016nxn,Bose:2017jvk,Chatziioannou:2017ixj}. Several features of the
signal are understood, but quantitative models are missing. We anticipate that
\core data will be used to develop new post-merger models to be employed
for the analysis of current and third-generation
detectors. The latter are the most promising observatories to capture
high-frequency GW signals, e.g.~\cite{Clark:2014wua,Radice:2016gym,Radice:2016rys}. 

Our data can also be injected in synthetic detector noise to test 
parameter estimation pipelines, similarly to what was done 
for BBHs~\cite{Abbott:2016wiq}.
For BNSs, however, complete waveforms spanning thousands of
GW cycles during the inspiral and tens of GW post-merger cycles would be needed. 
To address the problem, we generate hybrid waveforms combining 
analytical models and NR data and covering the frequency range
of ground-based interferometers \cite{Dudiinprep} (see also 
\cite{Read:2009yp,Read:2013zra,Hotokezaka:2016bzh}). 
We release 18 of these hybrids corresponding to equal, unequal masses
and spinning BNSs.

The \core database will have a reach beyond the applications we
have just discussed. In the future, we plan to include more quantities 
from our simulations. For example, mass
outflows ejected during merger, e.g.~\cite{Hotokezaka:2012ze,Bauswein:2013yna,
Sekiguchi:2015dma,Dietrich:2015iva,Radice:2016dwd,Dietrich:2016fpt,Lehner:2016lxy,
Fujibayashi:2017puw,Bovard:2017mvn}, and
disk masses and profiles \cite{Radice:2017lry}. These data will be crucial for 
the interpretation of EM counterparts.

\section*{Acknowledgments}
  We thank A.~Nagar and S.~Ossokine for discussions and A.~Sternbeck
  at TPI Jena for technical help.  

  T.D.\ acknowledges support by the European Union’s Horizon 
  2020 research and innovation program under grant
  agreement No 749145, BNSmergers.
  D.R.\ acknowledges support from a Frank and Peggy Taplin Membership at the
  Institute for Advanced Study and the
  Max-Planck/Princeton Center (MPPC) for Plasma Physics (NSF PHY-1523261).
  S.B.~acknowledges support by the European Union's 
  H2020 under ERC Starting Grant, grant
  agreement no.\ BinGraSp-714626. 
  A.P.\ acknowledges support from the INFN initiative 
  “High Performance data Network” funded by CIPE.  
  R.D.\ and B.B.\ were supported by DFG grant BR 2176/5-1.
  S.V.C.\ and R.D.\ were supported by the DFG Research Training 
  Group 1523/2 "Quantum and Gravitational Fields".
  W.T.~was supported by the National Science Foundation under grants
  PHY-1305387 and PHY-1707227.
  
  Computations were performed on the supercomputer SuperMUC at the LRZ
  (Munich) under the project number pr48pu, Jureca (J\"ulich) 
  under the project number HPO21, on the supercomputers
  Bridges, Comet, and Stampede (NSF XSEDE allocation TG-PHY160025), on
  NSF/NCSA Blue Waters (NSF PRAC ACI-1440083), on Marconi (PRACE proposal 2016153522 and
  ISCRA-B project number HP10B2PL6K), 
  on the Hydra and Draco clusters of the Max Planck Computing and Data Facility, 
  the compute cluster Minerva of the 
  Max-Planck Institute for Gravitational Physics, 
  and on the ARA cluster of the University of Jena. 

\bibliography{paper20180605.bbl}

\end{document}